\documentstyle[prl,aps,epsfig,floats]{revtex}

\newcommand\beq{\begin{equation}}
\newcommand\eeq{\end{equation}}
\newcommand\bea{\begin{eqnarray}}
\newcommand\eea{\end{eqnarray}}
\newcommand\nonum{\nonumber}
\newcommand\sqpi{{\sqrt \pi}}
\newcommand\tphi{\tilde\phi}
\newcommand\ua{\uparrow}
\newcommand\da{\downarrow}

\begin{document}

\draft

\textheight=23.8cm
\twocolumn[\hsize\textwidth\columnwidth\hsize\csname@twocolumnfalse\endcsname

\title{\Large Transport through quasi-ballistic quantum wires: the role
of contacts}
\author{\bf Siddhartha Lal$^1$, Sumathi Rao$^2$ and Diptiman Sen$^1$} 
\address{\it $^1$ Centre for Theoretical Studies,
Indian Institute of Science, Bangalore 560012, India \\ 
\vskip .5 true cm
$^2$ Harish-Chandra Research Institute, \\
Chhatnag Road, Jhusi, Allahabad 211019, India}

\date{\today}
\maketitle

\begin{abstract}
We model one-dimensional transport through each open channel of a quantum wire 
by a Luttinger liquid with three different interaction parameters for the 
leads, the contact regions and the wire, and with two barriers at the 
contacts. We show that this model explains several features of recent 
experiments, such as the flat conductance plateaux observed even at finite 
temperatures and for different lengths, and universal conductance corrections 
in different channels. We discuss the possibility of seeing resonance-like 
structures of a fully open channel at very low temperatures. 
\end{abstract}
\vskip .5 true cm

\vskip.5pc
]
\vskip .5 true cm


Recent advances in the fabrication of quantum wires within GaAs-AlGaAs
heterostructures have made it possible to study their electronic transport 
properties in great detail
\cite{tarucha,thomas,yacoby1,liang,krist,yacoby2,bkane,reilly}. 
These studies show some puzzling features for the conductance of quantum 
wires; specifically, the flat conductance plateaux lying {\it below} 
integer multiples of $g_0 \equiv 
2e^2 /h$ \cite{tarucha,yacoby1,liang,bkane,reilly}. At the same time, the 
theory of Tomonaga-Luttinger liquids (TLL) has led to a general framework for 
understanding the effects of impurities and finite temperature and wire length 
on the conductance of strongly correlated electron systems in one dimension
\cite{kane,safi,maslov,furusaki,safilong}. In this Letter, we 
propose a model for a quantum wire (QW) based on TLL theory which 
provides a unified way of understanding a large class of 
experiments including the features mentioned above.

Our model of a QW is motivated partly by the way these structures
are fabricated and partly by the experimental observations. The 
electrons are first confined to a two-dimensional region which 
is the inversion layer of a GaAs heterostructure.
Within that layer, a gate voltage $V_G$ is applied in a small region which 
further confines the electrons to a narrow constriction called the QW; this 
is typically a few microns long \cite{tarucha,yacoby1,liang}. Within the QW, 
the electrons feel a transverse confinement potential produced by $V_G$; this 
leads to the formation of discrete subbands or channels. 
As argued in Ref. \cite{chamon}, the electrons from the 
two-dimensional electron gas (2DEG) can enter or leave a one-dimensional
system like a wire only if they are in a zero angular momentum state with 
respect to the appropriate end of the wire. Since the radial coordinate is the 
only variable appearing in the wave function of such a state, the 
2DEG electrons which 
contribute to the conductance may be modeled as one-dimensional noninteracting 
Fermi liquid systems lying on the two sides of the QW \cite{safi}; we will 
refer to these two semi-infinite systems as the {\it leads}. However when the
electrons enter the constriction, they 
interact via the Coulomb repulsion. If the Coulomb repulsion is 
approximated by a short range interaction, the electrons in the wire can
be described by a TLL. In fact, each open channel (defined below)
can be modeled by a separate TLL. In addition, the charge and spin 
degrees of freedom are governed by independent TLL's if the interactions 
are invariant under spin rotations and there is no magnetic field.

The simplest model which incorporates all these features
is a one-dimensional system in which the TLL parameters (an interaction 
parameter $K$ and the quasiparticle velocity $v$) are functions of 
the coordinate $x$ as follows. If the QW lies in the range $0 < x < l$,
we let $(K,v)$ take the values $(K_L, v_L)$ for $x < 0$ and $x >l$ (the 
leads), and the values $(K_W, v_W)$ for $0 < x <l$ \cite{safi,maslov}. 
(The parameters $K_W$ and $v_W$ will also carry charge and spin labels 
$\rho$ and $\sigma$; we will include these later). Here $v_L$ is equal to 
the Fermi velocity $v_F ={\sqrt {2E_{F2D}/m}}$ of the electrons in 
the 2DEG (thus $v_L$ depends on the density and effective mass of 
the electrons in the 2DEG but {\it not} on any parameters of the QW), while 
$v_W$ is the velocity of the quasiparticle excitations inside the QW. 
Since the electrons are taken to be noninteracting in the 
leads, we set $K_L =1$; in the QW, repulsive interactions make 
$K_W < 1$. However, this model is not adequate for 
explaining the observed conductances. The main difficulty is that $v_W$ 
actually varies from channel to channel and depends on the gate voltage. 
The lowest energies $E_s$ in each channel are given by the discrete 
energy levels for the transverse confinement potential, and they can be 
shifted by changing the gate voltage $V_G$ \cite{buttiker}. Then $E_{F1D}$ in
the $s^{\rm th}$ channel is
defined to be $E_{F2D} - E_s$. If this is positive, the channel is open and
the electrons have a velocity $v_W (e) = {\sqrt {2E_{F1D}/m}}$
which is related to the velocity $v_F$ in the leads by $v_W (e) =
{\sqrt {v_F^2 - 2E_s /m}}$. Hence, the quasiparticle velocity $v_W$ 
also depends on the channel index $s$. However, the observed conductances 
show some features which are both channel and gate voltage independent. 

We will therefore consider a different 
model which additionally has two {\it contact regions} of length $d$ lying 
between the QW and the leads as shown in Fig. 1, so that the total length of
the QW system is $L=l+2d$. The TLL 
parameters in the contact regions are denoted by $(K_C ,v_C)$; these will
also have spin and charge labels as indicated below. It is important for us 
that $v_C$ should be independent of the gate voltage $V_G$ which is only 
felt within the QW; thus $v_C$ is a function only of $v_F$ and the 
interaction parameter $K_C$. This is physically reasonable if we visualize 
the contacts as regions where the gate voltage is not yet felt by the 
particles,  so that the Fermi velocity of the 
electrons has not changed from its value in the leads; however the electrons 
may begin to interact with each other in the contacts, so that $K_C$ could 
be less than $1$. In short, the contact regions model the fact that in 
many experiments, the electrons do not directly go from the 2DEG to the 
QW; there is often a smooth transition region between the two. In fact, 
a recent experiment has explicitly studied the effect of a transition region 
between the 2DEG and the wire, and has shown that a region of an
appreciable length of about $2-6 \mu m$ is required to cause 
backscattering \cite{yacoby2}. This makes our modeling of
the contact region as a Luttinger liquid of finite length $d$, 
rather than point-like, quite plausible. 

Note that the TLL's appearing in our model differ somewhat from a 
conventional TLL in which the electron velocity is related to the density. 
We are making the quasi-ballistic assumption that the electrons come in 
from the Fermi surface of the 2DEG, and shoot through the contact and 
wire regions where they interact with each other. Hence the density of the 
electrons in the contacts and wire do not play a direct role in our 
model; the quasiparticle velocity $v_W$ is determined 
primarily by $v_F$ and the subband energies in the QW. The idea that the
properties of the one-dimensional system are governed by $E_{F2D}$ has 
been used earlier in Ref. \cite{matveev} for a quantum point contact.

Given the Lagrangian density of a massless bosonic field in $1+1$ dimensions 
as \cite{kane} $${\cal L} (\phi ; K,v) = (1/2Kv) (\partial \phi /\partial 
t)^2 - (v/2K) (\partial \phi /\partial x)^2 ,$$ the bosonized action for the 
model described above is given by
\bea
S_0 &=& \int dt ~[ \int_{-\infty}^0 dx {\cal L}_1 + \int_{l+2d}^{\infty} 
dx {\cal L}_1 + \int_0^d dx {\cal L}_2 \nonumber \\
&& ~~~~~~~~~~ + \int_{l+d}^{l+2d} dx {\cal L}_2 + \int_d^{l+d} dx 
{\cal L}_3 ]~, 
\label{s0}
\eea
where
\bea
{\cal L}_1 ~&=&~ {\cal L} (\phi_\rho ; K_L , v_L ) ~+~ 
{\cal L} (\phi_\sigma ; K_L , v_L ) ~, \nonum \\
{\cal L}_2 ~&=&~ {\cal L} (\phi_\rho ; K_{C\rho} , v_{C\rho} ) ~+~ 
{\cal L} (\phi_\sigma ; K_{C\sigma} , v_{C\sigma} ) ~, \nonum \\
{\cal L}_3 ~&=&~ {\cal L} (\phi_\rho ; K_{W\rho} , v_{W\rho} ) ~
+~ {\cal L} (\phi_\sigma ; K_{W\sigma} , v_{W\sigma} ) ~,
\eea
where the charge and spin fields $\phi_\rho$ and $\phi_\sigma$ are continuous
at the points $x=0,d,l+d$ and $l+2d$. These fields are related to the bosonic
fields of the spin-up and spin-down electrons as $\phi_\rho = (\phi_\ua +
\phi_\da)/{\sqrt 2}$ and $\phi_\sigma = (\phi_\ua - \phi_\da )/{\sqrt 2}$.
 
In addition to the five different regions, our model includes two barriers.
The motivation for considering the barriers is 
two-fold. Since the geometry does not always change adiabatically 
from the 2DEG to the QW, one expects some scattering from the transition 
regions between the two \cite{yacimry}. Secondly, we have assumed that the 
strength of the electronic interactions change from zero in the 2DEG to a 
finite value in the contact regions; we will show elsewhere that this can 
produce some barrier-like scattering \cite{lal}. Although the scattering 
produced by the changes in geometry and interaction could, in principle, 
occur from anywhere in the contact regions, it is easier to study if we 
model it by $\delta$-function potentials placed at 
the junctions of the lead and contact regions, i.e., at the 
points $x=0$ and $x=l+2d$. These barriers take the form of spin-independent 
$\delta$-function potentials. Following Ref. \cite{safi2}, we can show 
that the results given below do not change if we consider extended barriers, 
as long as they lie entirely within the contact regions.

To summarize our model, the contact regions including the barriers are 
identical for all the subbands since the TLL parameters in the contacts only
depend on $E_{F2D}$. It is only inside the quantum wire that the TLL
parameters are different for different subbands.
Let us now consider what our model
yields for the conductance as a function of the various parameters. To begin 
with, we ignore the two $\delta$-function barriers and the gate voltage, and 
consider the action in Eq. (\ref{s0}). The conductance can be obtained 
from the frequency-dependent Green's function $G_\omega (x,x^\prime )$ which 
can be computed exactly \cite{maslov}. In the zero-frequency (dc) limit, we 
find that the conductance in each channel is given 
by $G = K_L g_0$ independent of
all the TLL parameters in the contact and QW regions. For $K_L =1$, this is
exactly the result expected for electrons with spin in the absence of any
scattering. This is at variance with the experimental observations which do 
show plateaux in the conductance, but at values which are renormalized down 
by a certain factor from the above values (see Fig. 2 of Ref. \cite{yacoby1}). 
It is notable that although the renormalization factor is sample dependent, 
it seems to be independent of the number of channels involved in the 
conductance \cite{yacoby1,liang,reilly}. 

The effect of the two
barriers is best studied using the effective action technique 
\cite{kane}. We first integrate out the bosonic fields at all points 
except the junctions at $x=0, d, l+d$ and $l+2d$; the fields at these
four points will be denoted by $\phi_{1a}, \phi_{2a}, \phi_{3a}$ and 
$\phi_{4a}$ respectively, where $a =\rho$ or $\sigma$. 
The expression for general frequency will be given elsewhere \cite{lal}. In 
the high-frequency limit $\omega \gg v_C /d$ and $v_W /l$, the action is 
given by
\bea
S_{eff} ~=~ \int \frac{d\omega}{2\pi} ~\vert \omega \vert ~[ && 
\frac{K_L + K_{C\rho}}{2 K_L K_{C\rho}} ~(\tphi_{1\rho}^2 ~+~ \tphi_{4 
\rho}^2 ) \nonumber \\
&& +~ \frac{K_{C\rho} + K_{W\rho}}{2 K_{C\rho} K_{W\rho}} ~
(\tphi_{2\rho}^2 ~+~ \tphi_{3\rho}^2 ) \nonum \\
&& + ~~{\rm similar ~terms ~with} ~\rho \rightarrow \sigma ~]~,
\eea
where the tilde's denote the Fourier transforms of the fields in time. We 
see that the four fields decouple at high frequencies or high temperatures;
in that limit, $\omega$ is related to $T$ by $\hbar \omega \sim k_B T$.

In the low-frequency limit $\omega \ll v_C /d$ and $v_W /l$, the action is 
given by
\bea
S_{eff} ~= && ~\int \frac{d\omega}{2\pi}  \frac{\vert \omega \vert}{2
K_L} ~(~\tphi_{1\rho}^2 + \tphi_{4\rho}^2 + \tphi_{1\sigma}^2 + \tphi_{4
\sigma}^2 ~) \nonum \\
&+& \int dt ~[~\frac{v_{C\rho}}{K_{C\rho}d} \{ (\phi_{1\rho} -\phi_{2
\rho})^2 + (\phi_{3\rho} - \phi_{4\rho})^2 \} \nonumber \\
&& ~~~~~~~~~ +~~ \frac{v_{W\rho}}{K_{ W\rho}l} ~(\phi_{2\rho} -\phi_{3\rho})^2 
\nonumber \\
&& ~~~~~~~~~ + ~~ {\rm similar ~terms ~with ~} ~\rho \rightarrow \sigma ~]~.
\label{lowfreq}
\eea

Next we introduce the $\delta$-function barriers at two points and the gate 
voltage in the QW region; they are given by $V[\delta (x) + \delta (
x-l-2d)]$ and $(eV_G /\sqpi) \int_d^{l+d} \partial \phi_\rho / \partial x$ 
respectively, where $e$ is the charge of an electron. This part of 
the action takes the form
\bea
&& S_{gate} ~+~ S_{barrier} \nonum \\
&& =~ \frac{eV_G}{\sqpi} ~\int dt ~[\phi_{3\rho} - \phi_{2\rho} ] \nonum \\
&& ~~ + \frac{V}{2\pi \alpha} \Sigma_i \int dt [ \cos (2 \sqpi 
\phi_{1i}) + \cos (2 \sqpi \phi_{4i} + \eta ) ],
\label{eta}
\eea
where $i$ is summed over $\ua ,\da$, $\alpha$ is a short distance cutoff,
and $\eta$ is given in terms of the wave numbers in the contact
regions and the QW as $\eta = 2 k_C d + k_W l$. After adding this action
to (\ref{lowfreq}) and integrating out the fields $\phi_{2a}$ and $\phi_{3a}$, 
we obtain the following low-frequency effective action in terms of the fields
$\tphi_\rho=(\tphi_{1\rho} + \tphi_{4\rho})/2$, 
$n_\rho = (\tphi_{4\rho} - \tphi_{1\rho})/\sqpi$, and $\tphi_\sigma$, 
$n_\sigma$ (defined similarly to their charge counterparts),
\bea
&& S_{eff} \nonumber \\
&& ~= \int \frac{d\omega}{2\pi} \frac{\vert \omega \vert}{K_L} 
[(\tphi_{\rho} - \frac{\eta}{2\sqpi})^2 + \frac{\pi}{4}(n_{\rho} - \frac{
\eta}{\pi})^2 + \tphi_{\sigma}^2 + \frac{\pi}{4} n_{\sigma}^2 ] \nonum \\
&& ~~ + \int dt ~[~\frac{U_\rho}{2} ~(n_{\rho} - n_{0\rho})^2 ~+~ \frac{U_{
\sigma}}{2} ~n_{\sigma}^2 \nonumber \\
&& ~~~~~~~+ \frac{V}{2\pi \alpha} \{ \cos(\sqpi \tphi_{\rho}) 
\cos (\sqpi\tphi_{\sigma}) \cos 
(\frac{\pi n_{\rho}}{2}) \cos (\frac{\pi n_\sigma}{2}) \nonum \\
&& ~~~~~~~~~~~~~~ + \sin (\sqpi\tphi_\rho)\sin(\sqpi\tphi_\sigma) 
\sin(\frac{\pi n_\rho}{2}) \sin(\frac{\pi n_\sigma}{2}) \} ]. \nonum \\
&&
\label{s1}
\eea
Here we have shifted the fields $\tphi_\rho$ and $n_\rho$ by factors 
proportional to $\eta$ to highlight the symmetries of the action 
coming from the barriers. $U_\rho = \pi\Lambda_{C\rho}\Lambda_{W\rho}/
(\Lambda_{C\rho}+2\Lambda_{W\rho})$ is an effective charging energy of the 
charges confined between the two barriers with $\Lambda_{C\rho}=
v_{C\rho}/(dK_{C\rho})$, $\Lambda_{W\rho}=v_{W\rho}/(lK_{W\rho})$, and 
$U_{\sigma}$, $\Lambda_{C\sigma}$ and $\Lambda_{W\sigma}$ are defined 
similarly with $\rho \rightarrow \sigma$. $n_{0\rho} = \eta /\pi - 
eV_G /(\pi^{3/2}\Lambda_{W\rho})$ is the average number of charges 
between the two barriers. As seen in Ref. \cite{kane}, such an effective 
action has a symmetry that if $n_{0\rho}$ is tuned to be an odd integer 
(using the gate voltage $V_G$), then there are two degenerate ground states. 
Tunneling between these two degenerate ground states in the weak barriers 
limit corresponds to a resonance in 
the transport of electrons through the system. For weak barriers $V \ll
U_\rho$ and $U_\sigma$, we can expand the terms involving $U_\rho , 
U_\sigma$ and $V$ in (\ref{s1}) around
$n_\rho = n_{0\rho}$ and $n_\sigma = 0$; this gives an effective barrier 
which vanishes if $n_{0\rho}$ is an odd integer. We thus require $\eta$ to 
be a constant plus an odd integer times $\pi$ for resonance.

We now do a renormalization group analysis to see how the barrier strengths 
scale with the length and temperature \cite{kane,furusaki,safilong} and
compute the conductance. We will work in the weak 
barrier regime (rather than the strong barrier or weak tunneling regime) as 
we believe that the two junction barrier strengths are weak; any 
renormalization of their strengths will also be small since the 
total length of the contacts and QW is small. We define $T_d =\hbar v_{C\rho}
/(k_B d) $ and $T_l=\hbar v_{W\rho}/(k_B l)$. If we assume that 
$d \ll l$ for simplicity, then $T_d \gg T_l$. The 
conductances to leading order in the barrier strengths 
are obtained in the limits where (i) $T_d \ll T $, (ii) $T_l\ll T \ll T_d$,  
and (iii) $T \ll T_l$. In the low temperature limit of (iii), the 
particles are phase coherent over the whole wire. At the intermediate
temperatures of (ii), they are coherent only over the contact region. In the
high temperature limit of (i), they are incoherent. 
The conductance  in regime (i) is given by 
\beq
g = g_0 K_L [ 1 - c_1 T^{2(K_{eff} - K_L)} (|V(0)|^2 + |V(l+2d)|^2) ]. 
\label{corrone}
\eeq
Here $c_1$ is a dimensionful constant containing factors of the velocity 
$v_{C\rho}$ and the cutoff $\alpha$ (but it 
is independent of all factors dependent on the gate voltage $V_G$), while
$K_{eff} = K_L K_{C\rho}/(K_L + K_{C\rho}) + K_L K_{C\sigma}/(K_L + 
K_{C\sigma})$.
At intermediate temperatures in regime (ii), it is given by 
\bea
g = g_0 K_L &&[ 1 - c_2 T_d^{2(K_{eff} - {\tilde K_{eff}})} 
T^{2({\tilde K_{eff}}-K_L)} \nonumber \\
&& \quad(|V(0)|^2 + |V(l+2d)|^2) ]. 
\label{corrtwo}
\eea
Here $c_2$ is a constant similar in nature to $c_1$, but it
can depend on $v_W$ and  hence 
is  not independent of the gate voltage $V_G$, while $\tilde K_{eff}$ is 
also dependent on interactions within the wire and is given as
${\tilde K_{eff}} = K_L K_{W\rho}/(K_L + K_{W\rho}) + K_L K_{W\sigma}/(K_L 
+ K_{W\sigma})$.
For low temperatures $T \ll T_l$, the conductance is
\bea
g = g_0 K_L &&[ 1 - c_2 T^{2(K_L - 1)} T_d^{2(K_eff -{\tilde K_{eff}})}
T_l^{2(K_eff - K_L)} \nonumber \\
&& \quad \quad |V(0) + V(l+2d)|^2 ].
\label{corrthree}
\eea
where the two barriers are now seen coherently.
Here again, $c_3$ is a constant similar in nature to $c_2$. (Similar
expressions can be derived if $T_d \ll T_l$, but the conclusions stated below
remain unchanged). As observed in several experiments, these conductance 
expressions reveal that as {\it either} the temperature $T$ is raised {\it 
or} the total length $L$ of the contacts and QW is decreased, the conductance 
corrections become smaller and the conductance approaches 
integer multiples of $g_0$ as expected \cite{tarucha,yacoby1}. Furthermore, we 
can see from these expressions that in the high temperature limit i.e., when
$T \gg T_d, ~T_l$, the conductance corrections are independent of the QW 
parameters. Hence, they are {\it independent} of the gate voltage $V_G$ and 
of all factors dependent on the channel index. Thus they yield 
renormalizations to the ideal values which are themselves plateau-like 
and uniform for all channels. Such corrections to 
the conductance explain some of the puzzling features observed in the 
experiments of Ref. \cite{yacoby1}. They have a fairly long contact 
region of $d \sim 2 - 6 \mu m$, which corresponds to $T_d \sim 
0.2 - 0.7 K$; this is much less than the temperature range shown in Fig. 3
of Ref. \cite{yacoby1}. Similar flat and uniform conductance corrections  
have been seen in the experiments of Refs. \cite{liang,reilly}; this seems to 
suggest that their experiments also include contact regions and $T \gg T_d$.
Interestingly, the low temperature corrections do depend on 
quantum wire parameters and consequently, on the gate voltage. Thus, 
a concrete prediction from this model is that one would fail to see flat 
plateaux in the conductance for $T \ll T_d$ indicating that the corrections 
are dependent on the gate voltage at lower temperatures. This has, in fact, 
been observed in a recent experiment (see Fig.3 in Ref.\cite{reilly}), where 
the conductances at $1K$ show flat and channel independent plateaux, but 
at $50mK$ are neither flat nor channel independent.

Finally, we observe that the existence of two weak barriers at the contacts 
could lead to the occurrence of resonances in regime (iii), where there is 
phase coherence over the entire wire and contact regions. Resonances can only 
occur  when $n_{0\rho}$ (defined earlier) is an odd integer, i.e., the phase 
\beq
\eta ~=~2k_C d + k_W l ~=~ (2n+1) \pi ~+~ \frac{eV_G}{\pi^{1/2} 
\Lambda_{W\rho}} ~.
\label{condition}
\eeq
Experimentally, $E_{F1D}$ and therefore $k_W$ is tuned by the gate voltage as 
one sweeps across all the states of the open channel, until the next channel
starts opening. If resonant transmission is possible at some energies, one 
would expect enhanced conductances when $k_W$ matches the condition given in 
Eq. (\ref{condition}). Such peaks in the conductance of an open channel may 
already have been seen at $T = 50 mK$ at conductances close to multiples 
of $g_0$ in Ref. \cite{reilly}. We expect these resonances caused by the
contact barriers to survive when the channels are moved laterally, unlike 
resonances which may be caused by impurities present inside the wire.
 
To summarize, we have presented a general model which can be applied to a 
large class of quantum wires. We will present elsewhere \cite{lal} the 
details of all our calculations as well as various extensions which are of 
experimental interest, such as the effects of impurities inside the QW, 
resonances occurring on the rise between two subbands, and a magnetic field 
where some additional features are observed.


\begin{figure}
\begin{center}
\epsfig{figure=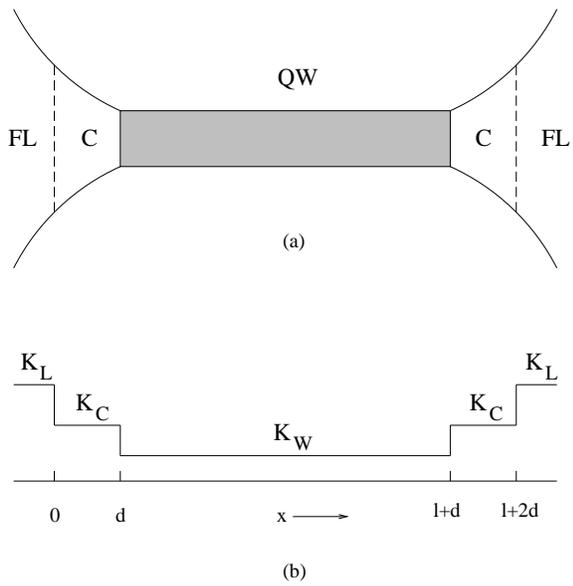,bbllx=50,bblly=0,bburx=550,bbury=800,height=12cm}
\end{center}
\vspace*{-2 cm}
\caption{Schematic diagram of the model showing the lead regions 
(marked FL for Fermi liquid), the contact regions (C) of length $d$, and the 
quantum wire (QW) of length $l$. The interaction parameters in these three 
regions are denoted by $K_L$, $K_C$ and $K_W$ respectively.}
\end{figure}

\end{document}